\documentclass[iop]{emulateapj}

\usepackage{graphicx}
\usepackage{natbib}
\usepackage{color}
\usepackage{comment}
\usepackage{upgreek}
\usepackage{epstopdf}
\usepackage{amsmath,amssymb}

\newcommand{\Pm}{\mr{Pm}}
\newcommand{\Reyn}{\mr{Re}}
\newcommand\bb[1]{\mbox{\boldmath{$#1$}}}
\newcommand\grad{\bb{\nabla}}
\newcommand\bcdot{\,\bb{\cdot}\,}
\newcommand\btimes{\,\bb{\times}\,}
\newcommand{\rmd}{\mr{d}}
\newcommand{\eb}{\bb{\hat{b}}}
\newcommand{\const}{\mr{const}}
\newcommand\bs[1]{\scriptscriptstyle\boldsymbol{#1}}
\newcommand{\bscdot}{\bs{\cdot}}
\newcommand{\bstimes}{\bs{\times}}
\newcommand{\bdbldot}{\,\bb{:}\,}
\newcommand{\betai}{\beta_{\mathrm{i}}}
\newcommand{\betaio}{\beta_{\mathrm{i}0}}
\newcommand{\betapar}{\beta_{\parallel\mathrm{i}}}

\newcommand{\pperp}{p_{\perp\mathrm{i}}}
\newcommand{\ppar}{p_{\parallel\mathrm{i}}}
\newcommand{\ROS}{\eb\eb\bdbldot\grad\bb{u}}
\newcommand{\mr}[1]{\ensuremath{\mathrm{#1}}}
\newcommand{\ba}[1]{\ensuremath{\langle #1 \rangle}}

\newcommand{\tcorr}{\ensuremath{t_\mr{corr}}}
\newcommand{\urms}{\ensuremath{u_\mr{rms}}}
\newcommand{\Brms}{\ensuremath{B_\mr{rms}}}
\newcommand{\nueff}{\ensuremath{\nu_\mr{eff}}}
\newcommand{\rhoi}{\ensuremath{\rho_\mr{i}}}
\newcommand{\Omegai}{\ensuremath{\Omega_\mr{i}}}
\newcommand{\vthi}{\ensuremath{v_\mr{thi}}}
\newcommand{\Deltai}{\ensuremath{\Delta_\mr{i}}}
\newcommand{\va}{\ensuremath{v_\mr{A}}}
\newcommand{\rhoio}{\ensuremath{\rho_\mr{i0}}}
\newcommand{\Omegaio}{\ensuremath{\Omega_\mr{i0}}}

\newcommand{\vao}{\ensuremath{v_\mr{A0}}}
\newcommand{\dio}{\ensuremath{d_\mr{i0}}}

\shorttitle{Collisionless Fluctuation Dynamo}
\shortauthors{St-Onge \& Kunz}

\begin{document}
\title{Fluctuation Dynamo in a Collisionless, Weakly Magnetized Plasma}
\author{Denis~A.~St-Onge}
\author{Matthew~W.~Kunz}
\affil{Department of Astrophysical Sciences, Princeton University, Peyton Hall, Princeton, NJ 08544, USA \\
Princeton Plasma Physics Laboratory, PO Box 451, Princeton, NJ 08543, USA}

\begin{abstract}
Results from a numerical study of fluctuation dynamo in a collisionless, weakly magnetized plasma are presented. The key difference between this dynamo and its magnetohydrodynamic (MHD) counterpart is the adiabatic production of magnetic-field-aligned pressure anisotropy by the amplification of a weak seed field. This in turn drives kinetic instabilities on the ion-Larmor scale---namely, firehose and mirror---which sever the adiabatic link between the thermal and magnetic pressures, thereby allowing the dynamo to proceed. After an initial phase of rapid growth driven by these instabilities, the magnetic energy grows exponentially and exhibits a $k^{3/2}$ spectrum that peaks near the resistive scale, similar to the large-magnetic-Prandtl-number ($\Pm\gg{1}$) MHD dynamo. The magnetic field self-organizes into a folded-sheet topology, with direction reversals at the resistive scale and field lines curved at the parallel scale of the flow. The effective $\Pm$ is determined by whether the ion-Larmor scale is above or below the field-reversing scale: in the former case, particles undergo Bohm-like diffusion; in the latter case, particles scatter primarily off firehose fluctuations residing at the ends of the magnetic folds, and the viscosity becomes anisotropic. The magnetic field ultimately saturates at dynamical strengths, with its spectral peak migrating towards larger scales. This feature, along with an anti-correlation of magnetic-field strength and field-line curvature and a gradual thinning of magnetic sheets into ribbons, resemble the saturated state of the large-$\Pm$ dynamo, the primary differences manifesting in firehose/mirror-unstable regions. These results have implications for magnetic-field growth in the weakly collisional intracluster medium of galaxy clusters.
\end{abstract}

\section{Introduction}\label{sec:intro}

The Universe is magnetized. While magnetic-field strengths of just ${\sim}10^{-18}~\rm{G}$ are required to achieve this both in our Galaxy and in clusters of galaxies,\footnote{This number is obtained by asking for what magnetic-field strength $B$ is the ion-Larmor radius $\rhoi\equiv\vthi/\Omegai$ roughly $1\%$ of some macroscale of interest $L$, where $\vthi\equiv(2T_\mr{i}/m_\mr{i})^{1/2}$ is the ion thermal speed, $\Omegai\equiv{eB/m_\mr{i}c}$ is the ion Larmor frequency, $T_\mr{i}$ is the ion temperature, and $m_\mr{i}$ is the ion mass. In the $T_\mr{i}\sim{5}~\mr{keV}$ intracluster medium (ICM), a typical macroscale is the thermal-pressure scale height, $L\sim{100}~\mr{kpc}$, and so $\rhoi\lesssim{0.01L}$ demands $B\gtrsim{10^{-18}}~\mr{G}$. This $B$ also ensures $\rho_\mr{i}\lesssim\lambda_\mr{mfp}$, the collisional mean free path. In the $T_\mr{i}\sim{0.5}~\mr{eV}$ interstellar medium, the same $B$ ensures $\rhoi\lesssim{0.01L}$ for $L\sim{1}~\mr{kpc}$.} observations of Faraday rotation, Zeeman splitting, and synchrotron emission all make the case for ubiquitous ${\sim}\mu\mr{G}$ fields \citep[e.g.,][]{Beck96,Carilli02,Beck15}. That these systems are not content with hosting weaker fields is surprising, at least until one realizes that the energy density of a ${\sim}\mu\mr{G}$ field is comparable to that of the observed turbulent motions; e.g., the {\em Hitomi}-observed velocity dispersion ${\approx}160~\mr{km~s}^{-1}$ in the ICM of Perseus \citep{Hitomi1} matches the Alfv\'{e}n speed $\va\equiv{B}/\sqrt{{4\pi}m_\mr{i}n}$ for the observed number density $n\approx{0.02}~\mr{cm}^{-3}$ if $B\approx{10}~\mu\mr{G}$. It is then natural to attribute the amplification and sustenance of (at least the random component of) the interstellar and intracluster magnetic fields to the fluctuation (or ``turbulent'') dynamo \citep{Batchelor,Zeldovich,ChildressGilbert}, by which a succession of random velocity shears stretches the field and leads on the average to its growth to dynamical strengths.

Despite being one of the most outstanding examples of magnetic self-organization and energy conversion in all of plasma physics, much about the fluctuation dynamo remains elusive \citep[e.g.,][]{BS05,KulsrudZweibel}. This is particularly true in weakly collisional plasmas, in which the collisional mean free path is comparable to or even larger than the macroscopic lengthscales of interest. In such systems, changes in magnetic-field strength lead to a field-oriented bias in the thermal motions of the particles, $\Delta\equiv{p_\perp/p_\parallel-1}$, where $p_\perp$ ($p_\parallel$) is the thermal pressure perpendicular (parallel) to the magnetic field \citep{CGL}. On large scales, the resulting pressure tensor induces an anisotropic response to the fluid flow \citep{Braginskii}, one which alters the efficacy of magnetic tension and promotes a folded magnetic geometry. On small scales, $\Delta$ serves as a source of free energy for rapidly growing kinetic-scale instabilities, namely firehose \citep{Rosenbluth56,Chandrasekhar58,Parker58,VedenovSagdeev58,Yoon93,HellingerMatsumoto00} and mirror \citep{Barnes66,Hasegawa69,SouthwoodKivelson93,Hellinger07}, whose growth and saturation impact the structure of the magnetic field and the effective plasma viscosity in a way controlled by the plasma beta parameter, $\beta\equiv{8\pi}nT/B^2$ \citep[e.g.,][]{Scheko_2008,SquireKunz17}.

These kinetic instabilities play a vital role in the plasma dynamo, as magnetic-field amplification is otherwise hampered both by phase mixing of the parallel rate of strain ($\ROS$, where $\eb\equiv\bb{B}/B$ is the magnetic-field unit vector and $\bb{u}$ is the fluid velocity) and by adiabatic invariance of $\mu\equiv{m}v^2_\perp/2B$. The latter sets stringent constraints on the allowable amount of magnetic-field growth \citep{helander_constraints}, while the former limits the velocity scales that can drive this growth. Fortunately, firehose and mirror saturate by developing sharp features in the magnetic field on ion-Larmor scales, which serve as either particle traps or instigators of pitch-angle scattering \citep{Kunz_kin,Riquelme_2015,Hellinger_2015}. These processes interrupt phase mixing and, in the case of scattering, break $\mu$-conservation.

In this {\em Letter}, we investigate how a weak seed magnetic field can be amplified to dynamical strengths in a turbulent, collisionless plasma, while allowing the plasma to respect bounds placed upon its pressure anisotropy by Larmor-scale kinetic instabilities. This work is complementary to studies of small-scale fluctuation dynamo in large-magnetic-Prandtl-number ($\Pm\gg{1}$) MHD fluids (e.g., \citealt{Scheko_theory,Scheko_sim}, hereafter S04; \citealt{Haugen04}, \S F).

%
%
\section{Method of Solution}\label{sec:method}

We present results from two numerical simulations of plasma dynamo using the second-order--accurate, hybrid-kinetic, particle-in-cell code \textsc{Pegasus} \citep{Pegasus}. The model equations governing the ion distribution function $f_\mr{i}(t,\bb{r},\bb{v})$ and the electromagnetic fields $\bb{E}(t,\bb{r})$ and $\bb{B}(t,\bb{r})$ are the kinetic Vlasov equation, Faraday's law of induction, and a generalized Ohm's law that assumes quasi-neutrality and includes the inductive and Hall electric fields, a thermo-electric field driven by pressure gradients in the massless electron fluid, and Ohmic ($\eta_\mr{Ohm}$) and fourth-order hyper ($\eta_\mr{H}$) magnetic resistivities (see equations (1)--(4) and (10) in \citealt{Pegasus}).

Both simulations are initialized with a stationary, spatially uniform, Maxwellian, ion-electron plasma in a triply periodic box of size $L^3$, threaded by a random, zero-net-flux magnetic field $\bb{B}_0$ with power at wavenumbers $kL/2\pi\in[1,2]$. The electrons are assumed isothermal with temperature $T_\mr{e}=T_\mr{i0}$, where $T_\mr{i0}$ is the initial ion temperature. Nearly incompressible turbulence is driven by applying a random, solenoidal, statistically non-helical force $\bb{F}(t,\bb{r})$ to the ions on the largest scales, $k_{F}L/2\pi\in[1,2]$. The amplitude of $\bb{F}$ is chosen such that the steady-state Mach number $M\equiv\urms/\vthi\sim{0.1}$, where $\urms$ is the rms ion flow speed. This amplitude is fixed; the amount of energy accepted by the plasma varies as its impedance self-consistently evolves. The forcing is time-correlated on $\tcorr\approx(k_F\urms)^{-1}$ using an Ornstein-Uhlenbeck process, which avoids spurious particle acceleration due to resonances with high-frequency power in, e.g., $\delta$-correlated driving \citep{Lynn12}.

The first simulation has $\betaio=10^6$ and $L/\rhoio=16$, and focuses on the early production of pressure anisotropy, its regulation by kinetic instabilities, the consequent generation of an effective collisionality, and the impact of these processes on magnetic-field amplification in the ``kinematic'' phase. It uses $504^3$ cells, $N_\mr{ppc}=216$ particles per cell, $\Omegaio\tcorr=16$, $\eta_\mr{Ohm}/\vao\dio=12.7$, and $\eta_\mr{H}/\vao\dio^3=13800$. The latter two parameters correspond to $\mr{Rm}_2\approx{3.2}\times10^4$ and $\mr{Rm}_4\approx{1.9}\times10^8$, where 
\begin{equation}
\mr{Rm}_{h}\equiv\frac{\urms}{k^{h-1}_{F}\eta_h}
\end{equation}
is a generalized magnetic Reynolds number for order-$h$ resistivity. The second run focuses on the ``nonlinear'' regime and the approach to saturation. It uses $\betaio=10^4$, $L/\rhoio=10$, $252^3$ cells, and $N_\mr{ppc}=216$. These parameters ensure that $\rhoi$ is well resolved, even in the saturated state in which $\betai{M^2}\sim{1}$ is anticipated. To maximize scale separation, only hyper-resistivity is used in this run, with $\eta_\mr{H}/\vao\dio^3=6$ ($\mr{Rm}_4\approx{1.1}\times{10^7}$). The viscosity, and thus the Reynolds number $\mr{Re}$, is determined self-consistently by wave-particle interactions and is not an input parameter as in the MHD dynamo. In both runs, the plasma starts well magnetized; a separate publication will focus on the transition between the unmagnetized and magnetized regimes. 

In what follows, $\langle\,\cdot\,\rangle$ ($\langle\,\cdot\,\rangle_\mr{p}$) denotes a box (particle) average.

%
%
\section{Results}\label{sec:results}

The plasma dynamo can be characterized by four distinct stages: (1) an initial period of fast, diffusion-free growth, during which ion-Larmor-scale firehose/mirror instabilities are excited; (2) a reduction in growth rate, leading to steady exponential growth similar to the kinematic regime of MHD dynamo; (3) a non-linear regime, in which the magnetic field becomes strong enough to influence the bulk plasma motion via the Lorentz force; and (4) the saturated state, in which the magnetic and kinetic energies become comparable. Results from both runs are used to elucidate each stage.

%
%
\begin{figure}
    \centering
    \includegraphics[width=0.45\textwidth]{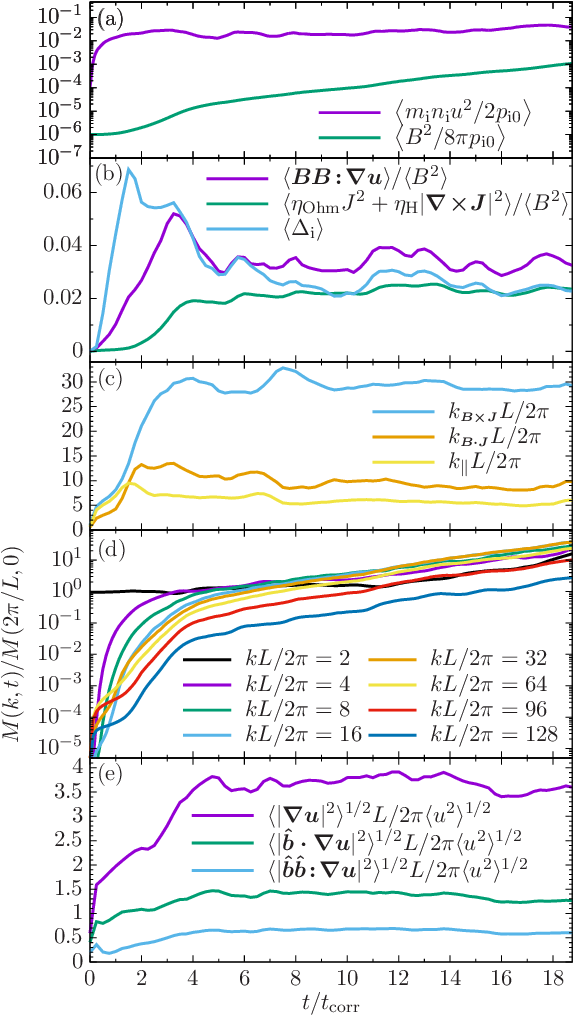}
    \caption{\label{beta6_energy} (a) Kinetic and magnetic energies; (b) parallel rate of strain, total magnetic dissipation, and pressure anisotropy; (c) characteristic parallel and perpendicular wavenumbers; (d) magnetic energy spectrum for select wavenumbers; and (e) components of the rate of strain, all for $\betaio=10^6$.}
\end{figure}

\subsection{Initial rapid-growth phase {\rm ($\betaio = 10^6$, $t/\tcorr \lesssim 5$)}}

\begin{figure*}
    \centering
    \includegraphics[width=0.95\textwidth]{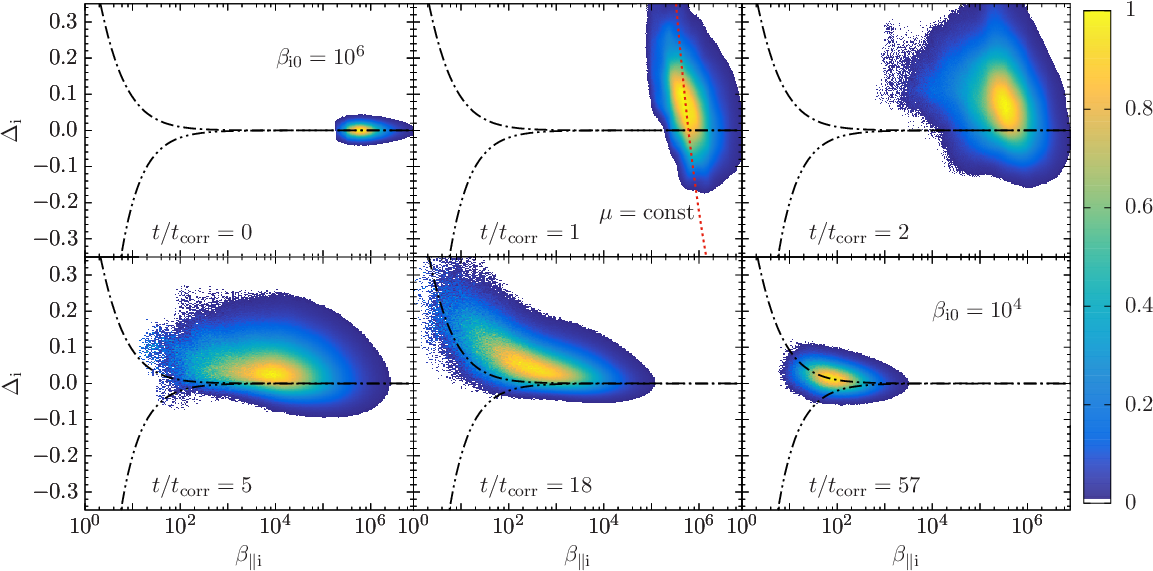}
    \caption{\label{aniso}Distribution of pressure anisotropy versus $\betapar$ in the rapid-growth ($t/\tcorr=0,1,2$) and kinematic ($t/\tcorr=5,18$) phases for $\betaio=10^6$, and in the saturated state ($t/\tcorr=57$) for $\betaio=10^4$. Dot-dashed (dot-dot-dashed) lines denote approximate mirror (firehose) instability thresholds. The red dotted line traces $\pperp/\ppar\propto\betapar^{-2}$, corresponding to evolution with $\mu=\const$.}
\end{figure*}

Figure~\ref{beta6_energy}(a) displays the box-averaged kinetic and magnetic energies versus time for the $\betaio=10^6$ run. The kinetic energy saturates within $t\approx\tcorr$ and a large-scale smooth flow is established. On the average, this flow amplifies the large-scale seed magnetic field, and rapid growth of magnetic energy occurs at $k\rhoi\approx{0.5}$--$1$ ($kL/2\pi\approx{4}$--$8$), adiabatically driving $\ba{\Deltai}>0$ (Figure~\ref{beta6_energy}(b); see also Figure~\ref{aniso}, $t/\tcorr=1$). Because $\betaio\gg{1}$, mirror instabilities are readily excited. From the standpoint of these mirror modes, the initial seed field ($kL/2\pi=1,2$) behaves as a local ``mean'' field on which they grow with oblique polarization $k_{\bs{B}\bstimes\bs{J}}>k_\parallel>k_{\bs{B}\bscdot\bs{J}}$ (Figure~\ref{beta6_energy}(c), $t/\tcorr\lesssim{1.5}$), where
\begin{gather*}
k_\parallel \equiv \left( \frac{\left\langle|\bb{B}\bcdot\grad\bb{B}|^2\right\rangle}{\langle B^4\rangle}\right)^{1/2} ,  \\*
k_{\bs{B}\bstimes\bs{J}} \equiv \left( \frac{\left\langle|\bb{B}\btimes\bb{J}|^2\right\rangle}{\langle B^4\rangle}\right)^{1/2}, ~ 
k_{\bs{B}\bscdot\bs{J}} \equiv \left( \frac{\left\langle|\bb{B}\bcdot\bb{J}|^2\right\rangle}{\langle B^4\rangle} \right)^{1/2}
\end{gather*}
are the characteristic wavenumbers of magnetic-field variation along ($k_\parallel$) and across ($k_{\bs{B}\bstimes\bs{J}},k_{\bs{B}\bscdot\bs{J}}$) itself and $\bb{J}=\grad\btimes\bb{B}$ is the current density (see \S3.2.1 of S04). Firehose-unstable modes are also triggered on ion-Larmor scales in regions of locally decreasing field and, in concert with mirror-unstable modes, ultimately generate sharp features in the magnetic field that trap and pitch-angle scatter particles. The latter produces an effective collisionality $\nueff$, which drives $\Deltai$ towards marginal stability (Figure~\ref{aniso}, $t/\tcorr={2,5}$) and ties the pressure anisotropy to the parallel rate of strain (Figure~\ref{beta6_energy}(b), $t/\tcorr\gtrsim{3}$). This leads to a Braginskii-like relation, $\Deltai\approx{3}\ROS/\nueff$, in which a balance obtains between adiabatic production and collisional relaxation, with $\nueff\lesssim\Omegai$.

%
%

%
%
\begin{figure*}
    \centering
    \includegraphics[height=0.7\textwidth]{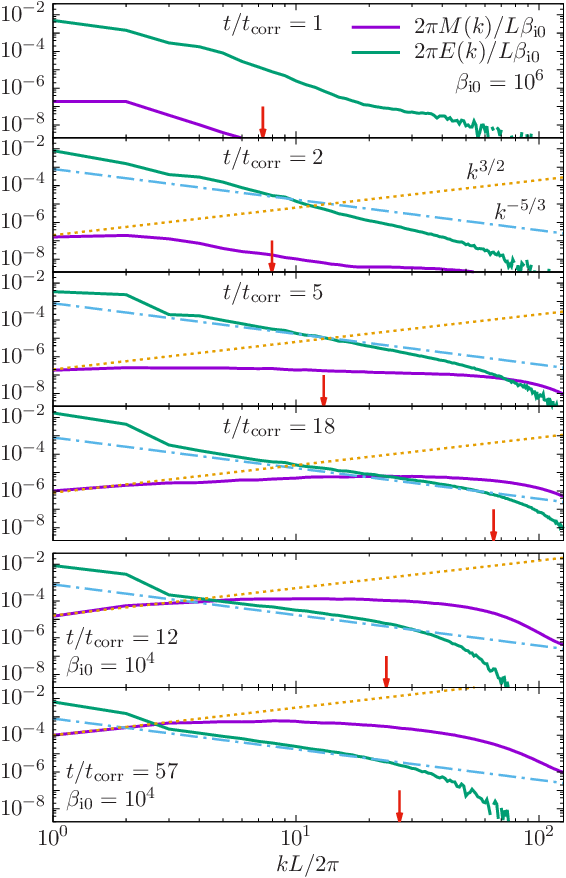}
    \qquad
    \includegraphics[height=0.7\textwidth]{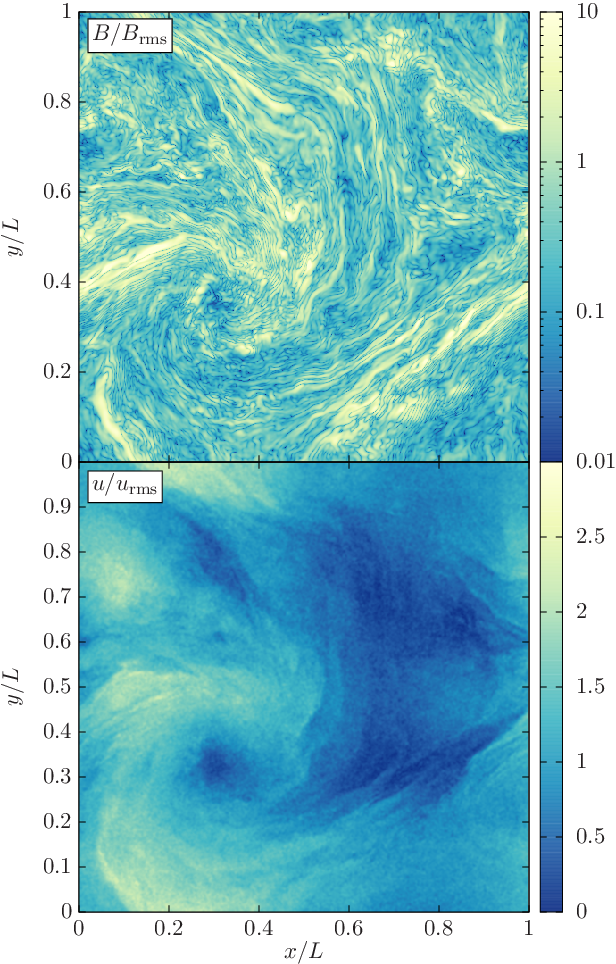}
    \caption{\label{spectra}{\it Left}: Magnetic- and kinetic-energy spectra for $\betaio=10^6$ ($t/\tcorr=1,2,5,18$) and $\betaio=10^4$ ($t/\tcorr=12,57$). Red arrows denote the wavenumber $\pi/\rho_\mr{median}$, where $\rho_\mr{median}$ is the median value of $v_\perp/\Omegai$. {\it Right}: Pseudo-color images of $B/\Brms$ and $u/\urms$ in a 2D slice during the kinematic phase for $\betaio=10^6$.}
    \label{energy_spec}
\end{figure*}

At the same time that the firehose and mirror instabilities saturate at $k\rhoi\lesssim{1}$ with $\delta{B}/B_0\sim{1}$, the magnetic field acquires energy at sub-ion-Larmor scales due to field-line stretching and folding by the large-scale flow (Figure~\ref{beta6_energy}(d), $t/\tcorr\gtrsim{5}$). The result is a much flatter angle-integrated magnetic-energy spectrum, $M(k)\equiv\frac{1}{2}\int\rmd\Omega_{\boldsymbol{k}}\,k^2\langle|\bb{B}(\bb{k})|^2\rangle$ (Figure~\ref{spectra}, $t/\tcorr=5$), than is seen in corresponding $\Pm\gg{1}$ MHD simulations. A change in the dominant magnetic-field topology accompanies this growth, with $k_{\bs{B}\bstimes\bs{J}}>k_{\bs{B}\bcdot\bs{J}}>k_\parallel$ indicating a folded geometry in which the field varies quickly (slowly) across (along) itself (Figure~\ref{beta6_energy}(c), $t/\tcorr\gtrsim{2}$).\footnote{The steady-state value of $k_\parallel$ in Figure~\ref{beta6_energy}(c) is an overestimate of the inverse fold length by a factor of $\approx$$2$--$3$, being biased towards larger $k_\parallel$ by ion-Larmor-scale firehose/mirror fluctuations.}

\subsection{``Kinematic'' phase {\rm ($\betaio = 10^6$, $ t/\tcorr \gtrsim 5$)}}
\label{sec:kinematic}

Eventually, this period of rapid growth ends. Figure~\ref{beta6_energy}(b) indicates that the reduction in growth rate is due to two effects. First, an appreciable fraction of the magnetic energy migrates to resistive scales, and magnetic diffusion becomes important. Secondly, $\ROS$ is sharply reduced between $t/\tcorr\approx{3}$--$5$, a feature we attribute to feedback from firehose/mirror fluctuations  \citep[e.g.,][]{Scheko_2008,Rosin_2011,Rincon_2015} and to parallel-viscous damping. Indeed, while $\ba{|\grad\bb{u}|^2}/\ba{u^2}$ increases substantially in that time interval, $\ba{|\ROS|^2}/\ba{u^2}$ remains nearly constant (Figure~\ref{beta6_energy}(e)); i.e., {\em the parallel rate of strain is suppressed}. That this is concurrent with the development of an angle-integrated kinetic-energy spectrum, $E(k)\equiv\frac{1}{2}\int\rmd\Omega_{\boldsymbol{k}}\,k^2\langle|\bb{u}(\bb{k})|^2\rangle$ (Figure~\ref{spectra}, $t/\tcorr=5,18$), that is \citet{Kolmogorov1941} (i.e., ${\propto}k^{-5/3}$) suggests that not all fluid motions cascade to the smallest scales; i.e., $\mr{Re}$ is {\it anisotropic}.

Thereafter, $\ba{B^2}$ grows exponentially (Figure~\ref{beta6_energy}(a), $t/\tcorr\gtrsim{5}$), much as in the kinematic-diffusive stage of the large-$\Pm$ MHD dynamo \citep[e.g.,][]{Scheko_theory}, with a growth rate $\gamma\doteq\rmd\ln\ba{B^2}^{1/2}/\rmd{t}=0.0093\Omegaio=0.15\tcorr^{-1}\approx\urms/L$ that becomes comparable at all scales (Figure~\ref{beta6_energy}(d), $t/\tcorr\gtrsim{5}$). The folded magnetic-field geometry previously established persists (Figure~\ref{beta6_energy}(c)), and $M(k)$ develops a \citet{Kazantsev} $k^{3/2}$ scaling with a peak near the resistive scale (Figure~\ref{energy_spec}, $t/\tcorr=18$). Such folded structure, accompanied by ion-Larmor-scale irregularities driven by firehose/mirror, is evident in the rightmost panels of Figure~\ref{energy_spec}, which display pseudo-color images of $B/\Brms$ and $u/\urms$ in a representative 2D slice. Suppression of parallel velocity variation is also apparent; while the turbulent velocity field is primarily large-scale, filamentary structures of near-constant $u$ develop along magnetic lines of force. Thus, there is a dynamical feedback of the magnetic field on the large-scale flow, even in the absence of a dynamically important Lorentz force, belying the ``kinematic'' moniker.

\begin{figure}
    \centering
    \includegraphics[width=0.45\textwidth]{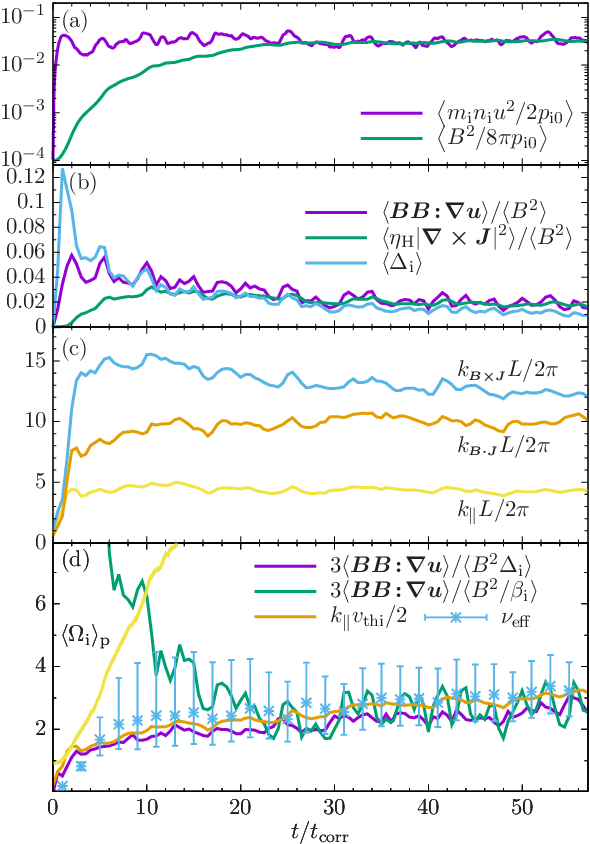}
    \caption{\label{beta4_energy} (a)--(c) As in Figure~\ref{beta6_energy}, but for $\betaio=10^4$. (d) Effective collision frequency (blue), compared to a ``Braginskii'' collision frequency (purple), the collision frequency required to maintain marginal firehose/mirror stability (green), a parallel-streaming frequency (orange), and the particle-averaged $\Omegai$ (yellow).}
\end{figure}

Because of the continuous energy injection and consequent magnetic-field amplification, along with insufficient scale separation between $L$ and $\rhoi$, exact marginal firehose/mirror stability cannot be maintained and a residual $\ba{\Deltai}\approx(0.02-0.03)\gg{1}/\betai$ persists for $t/\tcorr\gtrsim{5}$ (Figure~\ref{beta6_energy}(b)), with the bulk of the plasma approximately following the mirror threshold as $\betai$ decreases (Figure~\ref{aniso}, $t/\tcorr=18$). The regulation of $\Deltai$ is imperfect since, in order for saturated firehose/mirror instabilities to tightly regulate the pressure anisotropy near marginal stability, $\nueff\sim{S}\betai$ \citep{Kunz_kin,Melville}, where $S$ is the parallel rate of strain at the viscous scale (where it is largest). However, at $t/\tcorr=5$, $S/\Omegai\sim{10^{-2}}$ and $\betai\sim{10^5}$, thus requiring $\nueff\sim{10^3}\,\Omegai$ (!) Instead, $\nueff\ll\Omegai$ in the kinematic phase in both simulations, a point we have confirmed both indirectly, by comparing $\ROS$ and $\Deltai$ to infer $\nueff\approx{3}\ba{\bb{BB}\bdbldot\grad\bb{u}}/\ba{B^2\Deltai}$, and directly, by calculating the mean time over which $\mu$ changes for individually tracked particles (using the method described in \citet{Kunz_kin} and \citet{SquireKunz17}). The result is shown in Figure~\ref{beta4_energy}(d) for $\betaio=10^4$; qualitatively identical behavior is observed for $\betaio=10^6$.

There are two processes that contribute to $\nueff$, depending upon whether the majority of the particles' gyroradii is above or below the reversal scale of the magnetic field. In the former case, those particles sample several field-reversing folds during their gyromotion and thus undergo Bohm-like diffusion with $\nueff\sim\Omegai$. On the other hand, if the majority of particles have gyroradii below the field-reversal scale and remain well magnetized, then $\nueff$ is determined mainly by pitch-angle scattering off of firehose fluctuations, which populate regions of weak magnetic field where $\Deltai<0$.\footnote{The mirror instability only weakly scatters particles throughout much of its nonlinear evolution \citep{Kunz_kin,Melville}. Moreover, in turbulence where $S$ is a fluctuating quantity, the mirror instability is suppressed when $\betai>\Omega/S$ due to residual firehose fluctuations; see fig.~13 of \citet{Melville}.} As these regions occur primarily at the bends of the folded fields, we expect $\nueff\sim{k}_\parallel\vthi$, the inverse timescale for a thermal particle to traverse the length of a fold. Both of these contributions may be important, depending upon the structure of the magnetic field and the local magnetization of the plasma. In our runs, we witness only a brief moment in the evolution with $\nueff\sim\Omegai$, giving way to $\nueff\sim{k}_\parallel\vthi\ll\Omegai$ in the kinematic phase. It is only once $k_\parallel\vthi\sim{S}\betai$ that efficient regulation of $\Deltai$ is possible (\S\ref{sec:saturation}).

One consequence of $\nueff\ll\Omegai$ is an anisotropic viscosity, with Reynolds numbers $\mr{Re}\equiv\urms/(k_F\eta_\mr{visc})$ differing in the parallel and perpendicular directions: $\mr{Re}_\parallel\ll\mr{Re}_\perp$ \citep{Braginskii}. While the magnetic-field growth is controlled by $\mr{Re}_\parallel$ (since $\rmd\ln{B}/\rmd{t}\simeq\ROS\sim(\urms/L)\mr{Re}^{1/2}_\parallel$), the viscous cutoff $\ell_\mr{visc}$ seen in Figure~\ref{spectra} is determined by $\mr{Re}_\perp$ through the Kolmogorov relation $\ell_\mr{visc}\sim{L}\,\mr{Re}_\perp^{-3/4}$. Using classical transport theory to estimate the effective perpendicular ion viscosity $\eta_\mr{visc,\perp}\sim{0.1}\rhoi^2\nueff$, we find $L/\ell_\mr{visc}\sim(ML\Omegai/\rhoi\nueff)^{3/4}$. Taking $M$, $\Omegai$, $\rhoi$, and $\nueff$ from the run, we calculate a minimum value of $L/\ell_\mr{visc}\sim{10}$ at $t/\tcorr\approx{5}$, which grows exponentially to $L/\ell_\mr{visc}\sim{100}$ at $t/\tcorr\approx{18}$. This roughly agrees with the evolution shown in Figure~\ref{spectra}. Likewise, $\mr{Re}_\parallel$ can be calculated using the parallel viscosity for a magnetized plasma, $\eta_\mr{visc,\parallel}\sim\vthi^2/\nueff$. Once $\nueff\sim{k}_\parallel\vthi$, $\mr{Re}_\parallel\sim{M}(k_\parallel/k_F)\sim{1}$, suggestive of a $\mr{Pm}\gg{1}$ dynamo and consistent with the large drop in $E(k)$ at $kL/2\pi \approx 2$. Preliminary Braginskii-MHD dynamo simulations with $1\sim\mr{Re}_\parallel\ll\mr{Re}_\perp\sim\mr{Rm}$ and $-2/\betai\le\Deltai\le{1}/\betai$ enforced (e.g., following \citet{Sharma06} and \citet{Kunz_2012}) exhibit similar spectra and field-anisotropic flow to those presented here.

%
%

\subsection{Nonlinear regime and saturation $(\betaio = 10^4)$}
\label{sec:saturation}

Figure~\ref{beta4_energy}(a) shows the evolution of kinetic and magnetic energies for the $\betaio=10^4$ run. After evolving through the rapid-growth phase and a brief exponential kinematic phase, the field growth slows to become secular between $t/\tcorr \approx 12$--$24$ as the Lorentz force begins to affect the parallel-viscous-scale eddies (Figure~\ref{energy_spec}, $t/t_{\rm corr} = 12$; cf.~\citealt{maron04,cho09}). Saturation is ultimately reached with $\ba{B^2/4\pi}\approx\ba{m_\mr{i}nu^2}$ via a reduction of $\ROS$ (Figure~\ref{beta4_energy}(b), $t/\tcorr\gtrsim{25}$; S04).\footnote{The precise ratio of kinetic and magnetic energies in saturation may be influenced by the choice of Ohmic versus hyper resistivity. In a truly collisionless plasma, neither Ohmic nor hyper resistivity are guaranteed to be suitable replacements for electron-kinetic-scale physics. That being said, the resistive scale in the hot ICM, which we estimate following \citet{SchekoCowley06} using $\ell_\eta \sim L {\rm Rm}^{-1/2}$ with Spitzer (collisional) resistivity, is comparable to the present-day $\rhoi$, much larger than electron scales.} The ordering $k_{\bs{B}\bstimes\bs{J}}>k_{\bs{B}\bscdot\bs{J}}>k_\parallel$ established in the kinematic phase is preserved (Figure~\ref{beta4_energy}(c)), but the two perpendicular scales become closer to one another in saturation; i.e., the folded sheets evolve towards a ribbon-like structure, as seen in the $\Pm\gg 1$ MHD dynamo (S04). 

Despite the box-averaged equipartition between kinetic and magnetic energies, this balance is not scale-by-scale (Figure~\ref{spectra}, $t/\tcorr = 57$). Rather, there is an excess of the former at the forcing scales (since $E(k)\propto{k^{-5/3}}$) and an excess of the latter at smaller scales (since $M(k) \propto k^{3/2}$), although its peak has shifted towards smaller wavenumbers ($kL/2\pi \approx 5$--$10$) where the resistivity is negligible. It is because the folds exhibit spatial coherence at the flow scale that allows them to exert a back-reaction on the flow via the Lorentz force. Whether the shrinking gap between the parallel-viscous scale and the peak in $M(k)$ persists in higher-resolution simulations is of interest in the context of the intracluster magnetic field, whose spectrum is inferred to peak at scales (${\sim}1~\mr{kpc}$) far larger than the resistive scales \citep[e.g.,][]{VogtEnsslin}.

As in the $\betaio=10^6$ run, the pressure anisotropy becomes Braginskii-like, with $\ba{\Deltai}\propto\ba{\bb{BB}\bdbldot\grad\bb{u}}/\ba{B^2}>0$ (Figure~\ref{beta4_energy}(b), $t/\tcorr\gtrsim{5}$) and $\nueff\sim{k}_\parallel\vthi$ (Figure~\ref{beta4_energy}(d), $t/\tcorr\gtrsim{5}$). However, once $\betai$ decreases to ${\sim}50$ ($t/\tcorr\gtrsim{20}$), $\nueff\sim{S}\betai$ and $\Deltai$ is regulated close to the firehose/mirror thresholds (Figure~\ref{aniso}, $t/\tcorr=57$).

%
%
\begin{figure}
    \centering
    \includegraphics[width=0.45\textwidth]{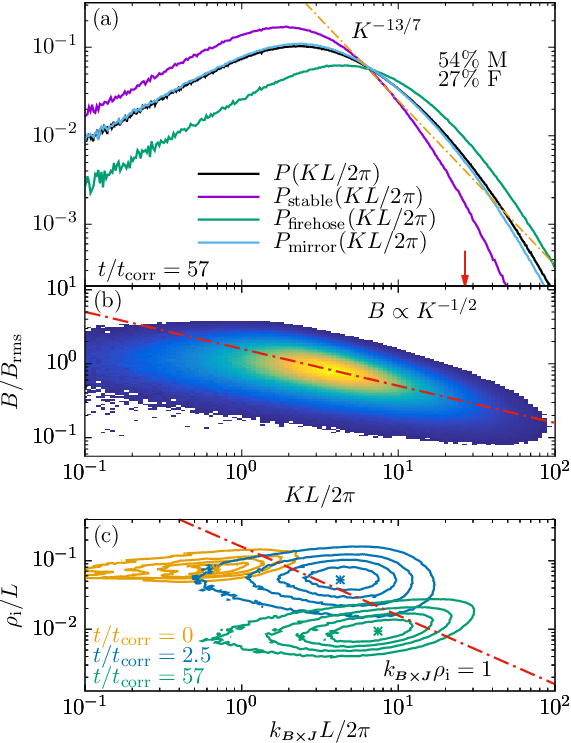}
    \caption{\label{curvature}(a) PDF of field-line curvature $K$ in saturation ($\betaio=10^4$, $t/\tcorr=57$) for firehose-unstable (green), mirror-unstable (blue), firehose/mirror-stable (purple), and all (black) regions. The predicted $K^{-13/7}$ scaling \citep{Scheko_theory2} is shown for comparison. The red arrow denotes the wavenumber $\pi/\rho_\mr{median}$. (b) Distribution of $K$ and $B$ in saturation. (c) Distribution of locally computed $\rhoi$ and $k_{\bs{B}\bstimes\bs{J}}$ for $\betaio=10^4$; contours are evenly spaced between $0.2$ and $1$.}
\end{figure}

Figure~\ref{curvature}(a) shows the probability distribution function $P(K)$ of the magnetic curvature $K\equiv|\eb\bcdot\grad\eb|$. In the MHD case, the tail of $P(K)$ relaxes to a $K^{-13/7}$ scaling \citep{Scheko_theory2} throughout the kinematic and saturated phases, depending only weakly on $\mr{Pm}$ (see fig.~25 of S04). While $P(K)$ in the plasma dynamo is peaked at similar values as those found in S04 ($KL/2\pi\approx{2}$), it is generally broader, and is dependent upon whether the host plasma is mirror unstable (blue; 54\% by volume), firehose unstable (green; 27\%), or stable (purple). Regions that are firehose unstable tend to have the largest curvature, for two reasons. First, $\Delta<0$ is generically produced in the stretched bends of the field lines, where $\rmd\ln{B}/\rmd{t}<0$ and $K$ is large. The reduction in effective field-line tension by $\Delta<0$ reinforces this trend. Secondly, firehose grows fastest at $k\rhoi\sim{1}$ and generates sharp kinks in the field lines on these scales. $K$ in mirror-unstable regions is also enhanced by the generation of mirror-shaped field lines. Despite this difference, there remains a strong anti-correlation between $B$ and $K$ in saturation (Figure~\ref{curvature}(b)), with $B\propto{K}^{-1/2}$ similar to the MHD case (cf.~fig.~17 of S04).

Finally, Figure~\ref{curvature}(c) displays the joint distribution of $\rhoi$ and $k_{\bs{B}\bstimes\bs{J}}$, each computed cell by cell, initially (orange), at the start of the kinematic phase (blue), and in saturation (green). Points rightward (leftward) of the dot-dashed line exhibit perpendicular magnetic structure on scales ${\lesssim}\rhoi$ (${\gtrsim}\rhoi$). At early times, this structure is driven by kinetic instabilities and the emergent folded-field geometry, with an appreciable fraction of the plasma having $\rhoi$ larger than the field-reversal scale. As $B$ increases, the mode of the distribution crosses into the magnetized region at $t/\tcorr\approx{5}$ and settles when the dynamo saturates ($t/\tcorr\approx{25}$). As this happens, the bulk of the plasma becomes well magnetized on the folding scale.

\section{Discussion}

The initialization and sustenance of the plasma dynamo rely heavily on the production and saturation of kinetic Larmor-scale instabilities, which effectively render the plasma weakly collisional by pitch-angle scattering particles. This scattering causes much of the overall evolution of the plasma dynamo to resemble the $\Reyn\sim{1}$, $\Pm\gg{1}$ MHD dynamo, including an analogous ``kinematic'' phase during which the magnetic energy experiences steady exponential growth. (Broad similarities between the MHD and collisionless fluctuation dynamo were also found by \citet{SantosLima} using a double-adiabatic fluid model with anomalous scattering to mimic the regulation of pressure anisotropy by firehose/mirror instabilities.) However, there are several differences, such as ion-Larmor-scale structure driven by firehose/mirror, a Kolmogorov-like cascade of perpendicular kinetic energy to ion-Larmor scales, and a field-biased anisotropization of the velocity field.

There is only one other publication to date using kinetic simulations to investigate the plasma dynamo \citep{Rincon_2016}.\footnote{A hybrid-kinetic study of dynamo in collisionless magnetorotational turbulence was presented in \citet{Kunz_mri}.} Those authors focused on the transition from the unmagnetized ($L/\rhoi\ll{1}$) to the magnetized ($L/\rhoi\gg{1}$) regime, with a parameter study conducted to obtain the critical $\mr{Rm}$ at which the dynamo operates. Where our results overlap with theirs, we find broad agreement. However, in the magnetized regime investigated here, computational expense prevented those authors from proceeding beyond the initial rapid-growth phase driven primarily by the mirror instability. Our finding that this rapid growth eventually gives way to a more prolonged and leisurely exponential growth casts doubt upon their suggestion that the plasma dynamo is self-accelerating, with $\gamma$ increasing as $B$ grows. Such an idea had been theorized previously: \citet{Scheko_expl}, \citet{Mogavero}, and \citet{Melville} conjectured that the firehose/mirror-endowed dependence of $\mr{Re}_\parallel$ on $\betai$ might result in an accelerating parallel-viscous-scale rate of strain, leading to explosive growth on cosmologically short timescales. However, \citet{Rincon_2016}'s finding of increasing $\gamma$ with decreasing $\betaio$ (${\propto}(k_F\rhoio)^{2}$ in their set-up) might instead be due to the role of $k_F\rhoio$ in setting $M$ for a given energy-injection rate and in facilitating initially rapid magnetic-field amplification by kinetic instabilities, topics that will be explored in a separate publication.

Clearly, efforts should focus on capturing the $\nueff\sim\Omegai\rightarrow{k}_\parallel\vthi\rightarrow{S}\betai$ transitions before saturation occurs at $\betai{M}^2\sim{1}$. Sorting this out is crucial not only for definitively testing theories of explosive dynamo, but also for determining the effective $\mr{Re}$ of the turbulent ICM \citep[e.g.,][]{Fabian05,ZuHone18}, which affects viscous heating \citep[e.g.,][]{Lyutikov07,Kunz_2011a,Zweibel18} and the integrity of cold fronts \citep[e.g.,][]{ZuHone15} and rising bubbles \citep[e.g.,][]{Fabian03}. Progress on these issues is now underway.

\begin{acknowledgments}
The authors are indebted to Alex Schekochihin, Steve Cowley, Francois Rincon, and Jono Squire for sharing with us their expertise on the small-scale dynamo, as well as the Wolfgang Pauli Institute in Vienna for its hospitality and support. This work was supported by U.S.~DOE contract DE-AC02-09CH11466, and made extensive use of the {\em Perseus} cluster at the PICSciE-OIT TIGRESS High Performance Computing Center and Visualization Laboratory at Princeton University. 
\end{acknowledgments}

\end{document}